# Gap Acceptance During Lane Changes by Large-Truck Drivers – An Image-based Analysis

Kazutoshi Nobukawa, Shan Bao, David J. LeBlanc, Ding Zhao, Huei Peng, and Christopher S. Pan

*Abstract*—This paper presents an analysis of rearward gap acceptance characteristics of drivers of large trucks in highway lane change scenarios. The range between the vehicles was inferred from camera images using the estimated lane width obtained from the lane tracking camera as the reference. Six hundred lane change events were acquired from a large-scale naturalistic driving data set. The kinematic variables from the image-based gap analysis were filtered by the weighted linear least-squares in order to extrapolate them at the lane change time. In addition, the time-to-collision and required deceleration were computed, and potential safety threshold values are provided. The resulting range and range rate distributions showed directional discrepancies, i.e., in left lane changes, large trucks are often slower than other vehicles in the target lane while they are usually faster in right lane changes. Video observations have confirmed that major motivations for changing lanes are different depending on the direction of move, i.e., moving to the left (faster) lane occurs due to a slower vehicle ahead or a merging vehicle on the right hand side, while right lane changes are frequently made to return to the original lane after passing.

## I. INTRODUCTION

The large-truck industry has been growing rapidly over the last few decades. There was about a 70% increase in the number of registered large trucks from the years 1975 to 2010, representing an increase of 3.8 million large trucks throughout the United States [1]. Correspondingly, large-truck safety research is becoming more and more important. In 2011, large trucks represented approximately 4% of all registered vehicles, but accounted for about 8% of all vehicles involved in fatal crashes [2]. According to the large-truck crash causation study (LTCCS) [3], only 17% of the total fatal large truck crashes were single-vehicle crashes, and about 48% were two-vehicle crashes that involved one large truck and one non-truck vehicle type.

Lane changes are one of the sources of major two-vehicle crashes that involve one large truck and one light vehicle. A lane change is defined as a maneuver that involves a deliberate and substantial shift in the lateral position of a vehicle when traveling in the same direction associated with simple lane changes, merge, exit, pass, and weave maneuvers [4]. Events when a vehicle moves onto the shoulder of the road or into an oncoming lane are not considered to be a lane change maneuver. The typical lane change crash scenario is that a vehicle changes lanes intentionally, and sideswipes or is sideswiped by another vehicle going straight in the target lane [5]. It was found that the number of lane change crashes that involved large trucks changing lanes and light vehicles going straight was twice as many as the number of crashes that involved large trucks going straight and light vehicles changing lanes [5]. Note that crashes that occur during large-truck lane changes are not necessarily a fault of the truck driver, since 60% of large trucks in multivehicle crashes are classified as having "no driver errors" [3].

According to a previous study based on the General Estimates System (GES) [6], 78% of lane change crashes occurred when the lane changing vehicle (subject vehicle or SV) and another vehicle in the target lane (principal other vehicle or POV) were traveling at closing speeds less than 15 mph (or 6.7 m/s), in which case the available gap could be very small, e.g., 11 feet (or 3.36 m) of gap for 0.5 s of the POV driver's reaction time, and 94% occurred with the closing speed less than 30 mph (or 13.4 m/s).

A gap, also referred to as a range, may be a primary safety measure for lane changes, defined by the difference in distance between the rear end of the subject vehicle and the front bumper of the POV in the target lane (Fig. 1). This term is also used to describe gaps in the context of merges into traffic or crossing streams of traffic [7, 8].

It is important to understand drivers' gap acceptance behavior when making lane changes for the purpose of truck driver safety benefit and future crash avoidance system design. However, existing studies are limited to passenger vehicle drivers. For example, a previous research project [9] conducted an on-road study with 16 participants and reported a mean rear gap of 30.7 m based on 109 lane change events. In [10], it was found that the average range was 46.7 m with 27% of lane changes occurring within 21.3 m of the preceding vehicle, for a total of 2,607 lane changes, and the average range rate (i.e., speed of the SV relative to the POV) was −1.25 m/s.

In reality, the decision making process of executing a lane

Manuscript received on XXXXXX. This work was supported by the National Institute for Occupational Safety Health (NIOSH) under Grant F031433.

K. Nobukawa (email: knobukaw@umich.edu), S. Bao (email: shanbao@umich.edu), and D. J. LeBlanc (leblanc@umich.edu) are with the University of Michigan Transportation Research Institute (UMTRI), Ann Arbor, MI 48109 USA.
H. Peng (email: hpeng@umich.edu) and D. Zhao (email: zhaoding@umich.edu) are with the Department of Mechanical Engineering, University of Michigan, Ann Arbor, MI 48109 USA.
Christopher S. Pan (email: syp4@CDC.gov) is with the National Institute for Occupational Safety Health (NIOSH), Morgantown, WV26505USA.

change happens a few seconds before the SV crosses the lane boundary. In this paper, however, we detect and present the gap at the exact moment when the SV crosses the lane boundary, because that time instant is precisely defined and can be reported without ambiguity. The exact moment when a driver assesses and decides to initiate a lane change is difficult to pinpoint.

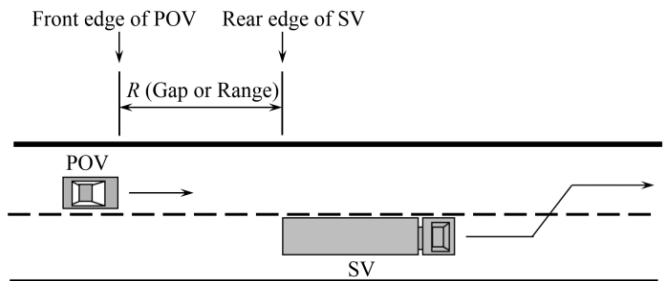

Fig. 1. Definition of gap in a lane change maneuver.

This paper focuses on the gap acceptance characteristics of truck drivers during lane changes in highway driving since previous studies are limited to passenger vehicles as mentioned above. The gap is estimated using an existing large-scale naturalistic driving database. Section II provides the descriptions of the naturalistic data, Section III explains details of the analysis, Section IV presents results of the gap analysis, and Section V provides concluding remarks.

## II. DESCRIPTIONS OF NATURALISTIC DRIVING DATA

Naturalistic driving data provide comprehensive information for analyses of driver behaviors for target scenarios without incurring artificial effects caused by predetermined driving conditions such as specified route and directed driving pattern. Typical conditions of a formal naturalistic driving data collection are that: 1) test subjects drive instrumented vehicles as their private or work vehicles used on a daily basis for a certain period of time, 2) test subjects drive the test vehicles free of guidance from the research personnel, and 3) test subjects are recruited from the general public or a relevant population group [11-13].

For the purpose of this research, the existing naturalistic driving data from the field operational test (FOT) for the study of the Integrated Vehicle-Based Safety Systems (IVBSS) [14] was used. The IVBSS data provide an exceptionally large amount of data with various information on multiple factors that are believed to contribute to motor vehicle crashes. Maintained by the University of Michigan Transportation Research Institute (UMTRI), the database is a repository and reporting mechanism for data obtained from naturalistic driving tests conducted on 16 light vehicles and 10 Class 8 tractors, with traveling distance of approximately 220,000 miles with light vehicles and 650,000 miles with the heavy vehicles. In the following, more detailed information for the large truck portion is provided.

The tractors were equipped with the integrated safety system, which includes a forward-collision-warning system (FCW), a lane-change/merge warning (LCM), and a lateral-drift warning system (LDW). The FCW system is designed to warn drivers of a potential rear-end crash with a lead vehicle while the LCM system alerts drivers of vehicles in the target lane. The LDW warning is issued when the vehicle deviates from the current driving lane without using turn signals. There were eight radars installed on each tractor to monitor surrounding on-road traffic [15]. Each truck was instrumented also to capture information regarding the driving environment, driver activity, system behavior, and vehicle kinematics, with a data collection frequency of 10 to 50 Hz. There are more than 500 data channels collected.

Eighteen commercial drivers from Con-way Freight participated in the IVBSS study to drive the trucks over a 10-month period. All drivers were required to have a minimum of two years of experience in driving commercial trucks. Due to the population of drivers available, all 18 drivers were males. The average age of the participants was 43 years old (range: 28 to 63 years old) with an average of 13 years of driving experience. They were instructed to drive naturally and were not explicitly encouraged to maintain safe headways. The test drivers used a driver-vehicle interface mounted on the dashboard to input the trip information such as the trailer length.

The first two months served as the baseline period during which warning functions were not presented to drivers, while the following eight months were the treatment period during which warnings functions were provided to drivers. During the baseline period, no system functionalities were provided to the drivers, but all sensors and equipment were running in the background. Although the test vehicles were equipped with an LCM warning system, it generated frequent false warnings due to reflection of the radar signal from non-target objects and thus the drivers were not in favor of the system according to the questionnaire. Therefore, the events from the baseline and treatment periods were combined in the gap analyses, assuming that the drivers primarily relied on their own gap judgement. Also, since the range of the rearward radar was short (33 m), an image-based technique was applied to rearview video images in order to estimate the gap.

## III. METHOD

### A. Overview

This section presents the analytical method of estimating the kinematic measures associated with the rearward gap at the time of lane changes of large trucks. The overall approach is described here, with sections following that address specific analysis elements. The gap analysis was conducted by an image-based method, and the results were evaluated using short-range rearward radars installed on the truck for proximate object detection. Manual identification of key image features was done on several images per lane change event, and models of camera imaging and gap dynamics were used to estimate these measures.

The estimation of the rearward gap, $R$, as shown in Fig. 1, was obtained based on the pinhole camera model, which assumes similar triangles to map the scene feature location onto the image position (Fig. 2),

$$Z_C = \frac{W}{w} z, \ (z = 1) \quad (1)$$

where $Z_C$ is the distance between the rearview camera and the front edge of the POV, which was mounted on the side mirror of the SV, and front end of the POV, $W$ is the real size of a reference feature, $w$ is the size of the feature in the normalized image coordinates (i.e., $z = 1$), which is transformed from the original image coordinates on the camera retina in pixels using the camera parameters. It is noted that the distance between the rearview camera and the rear edge of the trailer, $L$, needs to be subtracted from $Z_C$ to obtain the range, $R$.

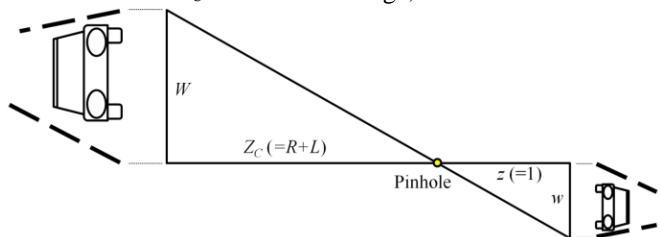

Fig. 2. Schematic of the pinhole camera model.

Selecting an appropriate reference feature is crucial for accurate results. Three options were initially considered: splay angle, POV width, and lane width at the POV location. The splay angle method uses the vertical pixel coordinate of the object in the image, and it only requires the camera height to be known [16]. However, this is very sensitive to a slight change of pitch angle of the SV, i.e., one degree of pitch angle error, which is typical for the trucks in highway driving, can cause an error of 60% for a distance of 50 m. Alternatively, using the vehicle width as a reference measurement provides a robust estimate since the relative distance of two points to specify the side edges of the POV on image will not change greatly under pitching disturbances. An issue with this option is that the vehicle width cannot be known easily since the model and year of each POV needs to be identified. On the other hand, the lane width at the POV location, which is the adopted reference measurement in this study, has the following advantages: 1) the lane width estimated by the lane tracking camera which was installed on the front edge of the SV is readily available, 2) lane width defined by the distance between two reference points on the image is unaffected by pitch disturbances or camera misalignment unlike the splay angle method, and 3) the lane width provides more pixels than the vehicle width.

Since the sampling rates were different between the onboard data acquisition system (10 Hz) and rearview camera (2 Hz), the range at the lane change time had to be inferred from the available data series. To do this, the series of range estimates in each lane change event was smoothed by the weighted linear least-squares technique and extrapolated – instead of interpolation due to an occlusion of lane marker by the truck body after the lane change time – by using the range rate estimated from the smoothed range estimates.

In the following, the analysis procedure is explained more in detail.

B. *Selection of lane change events*

In this paper, a lane change event is defined between the times when the center of the truck body reaches 0.1 m from the center of the original lane for the last time before crossing the lane marker and when the distance between the center of the truck body reaches 0.1 m from the center of the new lane for the first time after the whole truck body has moved to the adjacent lane. The lane change time was determined by the lane-tracking system (AssistWare SafeTrac2) when a significant jump in the lane-offset value was detected, which occurred when the inner side of the vehicle body was about to cross the lane marker. In this data set, the lane tracker flags lane changes and post-processing identifies which flags are associated with fully-executed lane changes.

In the IVBSS data, lane width is available for the current SV lane only since it was estimated by the lane tracking camera, which is a part of the lane departure warning system monitoring the forward view. Therefore, the lane width of the adjacent lane (i.e., target lane) at the POV location at the lane change time is not known directly. In this case, we assume that the target lane has a constant lane width, and finding suitable lane change events is crucial for the accuracy of the analysis results. An assumption in the event screening process is that through lanes on highways have a fairly constant lane width except for those adjacent to a ramp lane – the lane marker to separate these lanes is often missing and only outer lane markers of these lanes are available. In this case, the lane tracking camera would detect them as a single lane with a variable lane width.

As shown in Table I, the original set (A) of all the lane change events was reduced by applying various screening conditions. In the first screening, three conditions were applied: 1) highway (speed at least 55 mph (or 24.6 m/s)), 2) straight road (heading change within ±5 degrees), and 3) daytime (solar zenith angle between 0 and 96 degrees, or civil dusk). Using the resulting set (B), two subsets for non-ramp and ramp regions were created. Here, a ramp region was defined within a 500 m radius from the closest intersection point between the ramp lane and through lane obtained from a ramp location database, and if any portion of the vehicle trajectory during the time period between 2 s before the event start time and 5 s after the end time was inside this region, such a lane change event was classified into the ramp event set. Initially, the only non-ramp event set (C), in which lane change trajectories were outside the 500 m range, was considered since it is typical that the lane width is constant in this region, but only 31 left lane changes (and 280 right lane changes – See Table I) were detected with a clear POV appearance in the videos. Therefore, the ramp events were also analyzed to complement the non-ramp event set. The ramp event set (D) was created for four combinations of lane marker types, i.e., solid-dashed-solid, dashed-dashed-solid, solid-dashed-dashed, and all dashed (in the order of left to right regardless of the lane change direction). In this case, the target lane in right lane change events may have a variable lane width due to the ramp lane, and such events were eliminated during the manual feature selection explained in Section III.D.2).

TABLE I
REDUCTION OF THE EVENT SET

| Event set | Lane change direction | | Total |
|---|---|---|---|
| | Left | Right | |
| All lane change events (A) | 111,850 | 86,282 | 198,132 |
| High speed, straight, daytime (B) | 8,086 | 9,020 | 17,106 |
| Non-ramp events (C) | 727 | 809 | 1,536 |
| With video | 711 | 760 | 1,471 |
| With POV | 173 | 360 | 533 |
| Analyzed (E) | **31** | **280** | 311 |
| Ramp events (D) | 3,590 | 3,763 | 7,353 |
| With video | 3,537 | 2,606 | 6,143 |
| With POV | 1,057 | 1,561 | 2,618 |
| Analyzed (F) | **269** | **20** | 289 |
| Total events analyzed (E)+(F) | 300 | 300 | 600 |

From these two sets of lane change events, (C) and (D), a total of 600 lane change events (300 for each direction) were selected for the gap analysis. For the non-ramp events, all valid events (C) were explored, and the test set (E) was created, while the ramp events were randomly selected from the set (D) until the required number of events with a POV was collected to form the other test set (F).

There is a size difference between non-ramp event set (C) and ramp event set (D). The ramp event set is more than four time larger than the non-ramp event set. Although the threshold for the distance from ramp to separate the two types of lane change events was selected rather arbitrarily, considering the distance traveled in non-ramp region is longer than that in ramp region, this distinction may be related to motivating conditions for lane changes, such as other vehicles entering and exiting highway.

### C. Camera calibration

Since the pinhole camera model in (1) assumed the normalized image plane in which the coordinates are rectilinear, a camera calibration provided the camera parameters necessary to transform the data points in the original distorted image coordinates into the normalized image coordinates. There are two types of camera parameters, intrinsic and extrinsic parameters. The intrinsic parameters are associated with the characteristics of the camera itself such as the focal length, skewness, principal point, and distortion coefficients, while the extrinsic parameters account for the position and orientation of the camera.

Since the distance to the object was described with respect to the camera coordinate system and, as mentioned earlier in section III, the location of the POV appearing in the image will not affect the accuracy of the distance estimation, only the intrinsic parameters were necessary in the analysis. This eliminated the process of extrinsic parameter estimation for each subject vehicle and concern about the error due to a potential misalignment caused by vibrations and shocks over the test period.

The Camera Calibration Toolbox for MATLAB® [17] was used to obtain the camera parameters from the IVBSS rearview camera (PC88WR from Supercircuits, 30 Hz of frame rate with 2 Hz of image capture rate to save the storage space). It only requires photographs of a checkerboard in multiple orientations as inputs. A checkerboard with 3 cm by 3 cm squares containing 6 by 10 squares (i.e., 77 grid points) was prepared and 25 snapshots with different orientations were taken. The estimated focal length was 33 mm. The root-mean-square error of the reprojected grid points is about 0.75 pixels, which corresponds to approximately 1.2 % of error in estimating distance of a POV that is 30 m away from the camera.

### D. Gap estimation for single images

#### 1) Camera coordinate transformations

The equations used for the coordinate transformation from the original pixel coordinates obtained from the video images into the normalized coordinates [17] are summarized below.

The feature position in the world coordinates with respect to the reference frame attached to the center of the camera lens is

$$\mathbf{X}_C = \begin{Bmatrix} X_C \\ Y_C \\ Z_C \end{Bmatrix} \qquad (2)$$

where $X_C$ and $Y_C$ are the horizontal and vertical coordinates and the $Z_C$-axis coincides with the optical axis of the camera lens. The projection onto the normalized image plane (i.e., unity distance between the normalized image plane and pinhole) is

$$\mathbf{p}_n = \begin{Bmatrix} X_C/Z_C \\ Y_C/Z_C \\ Z_C/Z_C \end{Bmatrix} = \begin{Bmatrix} x \\ y \\ 1 \end{Bmatrix} \qquad (3)$$

where $x$ and $y$ are the horizontal and vertical image locations of the feature in the normalized coordinates. The transformation from the actual pixel coordinates on test images to the normalized coordinates is achieved by

$$\mathbf{p}_n = f^{-1}(\mathbf{p}_d) = f^{-1}(\mathbf{K}^{-1}\mathbf{p}) \qquad (4)$$

where $f(.)$ is the nonlinear transformation from $\mathbf{p}_n$ to the distorted normalized coordinates $\mathbf{p}_d$, $\mathbf{K}$ is the 3-by-3 camera matrix containing intrinsic camera parameters, and $\mathbf{p}$ is the actual image coordinates in pixels. (Refer to [17] for details.)

The idea is that the straight lane markers are also straight on the normalized image plane. In practice, two points found on the left and right lane markers ($\mathbf{p}^{left}$ and $\mathbf{p}^{right}$) were transformed into $\mathbf{p}_n^{left}$ and $\mathbf{p}_n^{right}$, and $w$ in (1) was computed by

$$w = \left| \mathbf{p}_n^{left} - \mathbf{p}_n^{right} \right|. \qquad (5)$$

#### 2) Range estimation for individual images

For each image, the feature points were manually selected (rather than automatically for purposes of robustness) from the original distorted image at two arbitrary points on the lane markers on each side of the adjacent lane, and at one point on the image at the bottom edge of the shadow under the POV (Fig. 3(a)). These points were then transformed by nonlinear transformation in (4) into the normalized image coordinates (Fig. 3(b)). Since the distorted lane markers in the video images become straight on the normalized image plane for a straight road, each lane marker was reconstructed by a line segment passing through the relevant points (Fig. 3(c)). Moreover, by assuming that the camera had been mounted on the truck with a small rotational angle about the lens axis, the horizontal segment passing the POV position drawn between the reconstructed lane markers represented the width $w$ in (1). Finally, the range estimation was achieved after subtracting the trailer length. Fig. 3(d) shows reconstructed lane markers and a horizontal line on the original image by re-projecting the line segments on the normalized plane.

At least seven consecutive video frames were used for the least squares model, but the number of frames with a good image quality was not known in advance. Therefore, the process of the range estimation started at the last available video frame before the lane change time and succeeded backward in time, and an event was discarded if the number of qualified frames was less than seven.

*3) Comparison between distance estimation results and radar data*

The accuracy of the image-based gap analysis was evaluated by comparing its results with data from the rearward radars. Since the detection range of the radar was 33 m, there were 15 lane change events containing the radar data, and fifty images were available with a POV in view.

The results show similar statistics for both the radar data and image-based analysis. The mean and standard deviation of the error between these sources are −4.84% and 6.03%, respectively.

It is noted that the distances compared here were measured between the devices and the POV (both camera and radar were mounted on the side mirror), instead of the distance between the rear edge of the truck and POV, in order to avoid introducing uncertainty from the variable trailer length. Also, only a single parameter set from a particular camera was used for the analyses for all trucks and for both sides since 1) cameras had been removed from the trucks, 2) not all cameras were available for the camera calibration, and 3) there were not significant variations between the model parameter sets for five different cameras that were investigated.

As shown in Fig. 4, the estimation error contains a bias in the negative direction, or the image-based gap analysis consistently underestimated the actual distance.

Table II shows summary information of the results.

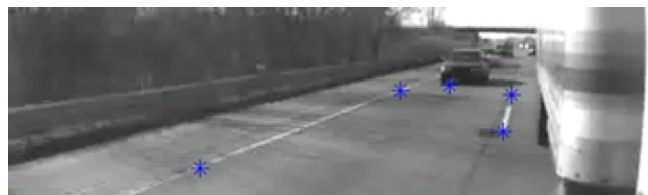
(a)

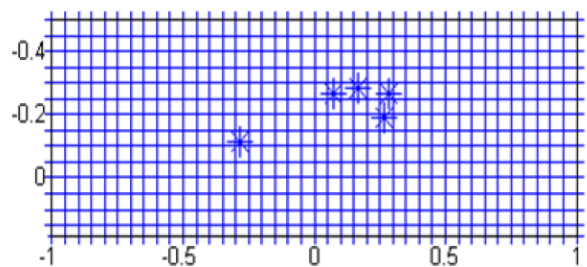
(b)

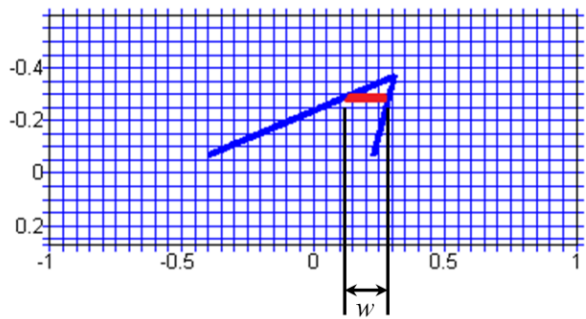
(c)

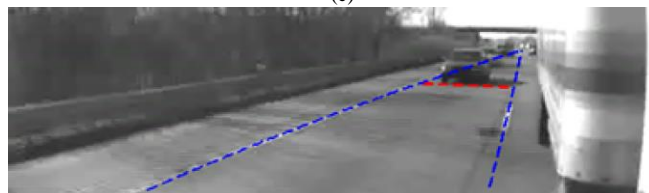
(d)

Fig. 3. Distance estimation process: (a) selected feature points on the original distorted image, (b) feature points in normalized coordinates, (c) reconstructed lane markers by linear extrapolation with a horizontal segment representing the lane width at the POV location, and (d) reprojected lane markers and POV position onto the original image.

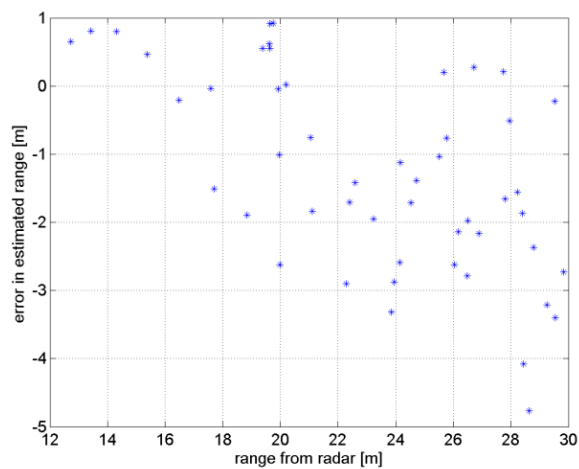

Fig. 4. Range estimation error vs. range from radar.

TABLE II
COMPARISON OF POV DISTANCE BETWEEN RADAR AND IMAGE-BASED ESTIMATES

| Distance from camera Mean [m] | Distance from radar Mean [m] | Error Mean [m] | Std [m] |
|---|---|---|---|
| 21.98 | 23.25 | −1.27 (−4.84 %) | 1.45 (6.03 %) |

*E. Least squares estimate for predicting the POV distance at lane change time using multiple images*

The obtained range estimates contained relatively large fluctuations mainly due to the relatively low resolution of the video images, and the range rate from the numerical differentiation was noisy as well. On the other hand, when a video frame was not available at the lane change time, the trajectory was extrapolated from the last available range estimate before that time to obtain a projected range estimate at that time, in which case an accurate range rate is required.

Here, the weighted first order linear least-squares technique was used to filter the series of range estimates, since the range rate does not change greatly in the lane change events in normal driving. In other words, the relative acceleration was small unless, for example, the POV responded by emergency braking, and therefore the range curve became approximately linear, and the range rate was approximately constant. Since the accuracy of the estimated distance is inversely proportional to the actual distance of the POV given a constant pixel error, the larger weight on the residual was applied to the closer POV. The weight is defined as

$$w_i = R_{min} / R_i \qquad (6)$$

where $R_{min}$ is the shortest range among the series of ranges available and $R_i$ is the range of the POV in the $i$-th frame. With the first order polynomial model, the range is represented by

$$R(t) = a_1 t + a_2 \qquad (7)$$

where $a_1$ represents the range rate (i.e., $\dot{R}(t) = a_1$) which is used to extrapolate the range curve at the lane change time by

$$R(t_{LC}) = a_1 \Delta t + R(t_n) \qquad (8)$$

where $\Delta t$ is the time period between the lane change time, $t_{LC}$, and the time of the last available frame, $t_n$.

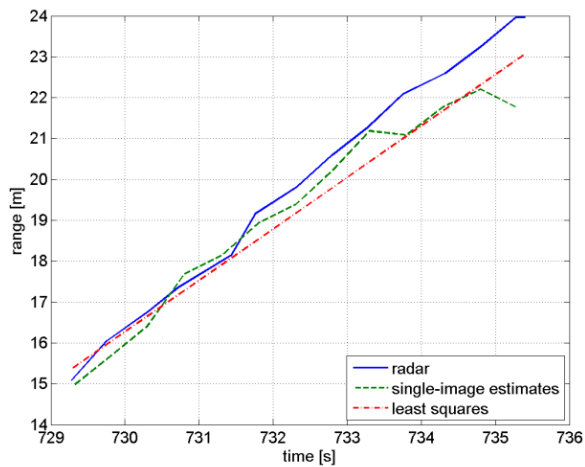

Fig. 5. Weighted least squares fit to improve range rate from camera data for an example event.

Fig. 5 shows the filtered result from the least squares for a single event, compared against the original series of range estimates obtained from the image-based analysis and corresponding range data from the radar which is assumed to be ground truth.

## IV. RESULTS AND DISCUSSION

Using the information from Table I, the POV appearance rate, defined by the ratio of the number of events with video available in which a POV exists to that of all the events also with video, was computed for each direction for both of the non-ramp and ramp event sets, i.e., 29.0% for left lane changes and 57.1% for right lane changes. The obtained results show that the appearance rate in the left lane changes is about half of that in the right lane changes. A possible reason for this difference is explained as follows. Usually large trucks are slower than passenger vehicles and stay in the rightmost lane on highways. In fact, the fleet that participated in IVBSS – Con-way Freight – had governors on their vehicles at the time of testing, limiting the truck speed to 62 mph (or 27.7 m/s). Therefore, when a truck changed lanes to the left lane to overtake a slower vehicle ahead or to yield the lane to another vehicle entering from a ramp, it returned to the original lane as soon as it completed overtaking a slower vehicle or a space became available in the original lane. In this case, the location or existence of the POV in the left lane is unrelated to the location of the slower vehicle ahead or the merging vehicle, but there would almost always be one present when the truck returned to its original lane, with a lane change to the right. This was investigated by classifying the lane change types by means of observing the videos. Table III shows the classification of lane change by scenario types for 142 events involving POVs found in the range of time-to-collision (TTC) [18, 19] (see (9) for its definition) between −10 and 10 s. Among the 142 sample events, 128 events (90.1%) are associated with the scenario of either overtaking slower vehicles (60.1%) or avoiding merging vehicles entering through ramps (30.0%). The numbers of left lane changes in the first and second scenarios are similar, but those of right lane changes are very different. This occurred because the vehicle that made the truck drivers change lanes was overtaken or became the POV more frequently in the first scenario than in the second. In the second scenario, the merging vehicle usually reached a sufficiently fast speed so as to enter the through lane after overtaking the truck.

TABLE III
NUMBER OF LANE CHANGES FOR TTC BETWEEN −10 AND 10 S FOR ASSOCIATED SCENARIOS

| Scenario | Lane Change Direction | | Total |
|---|---|---|---|
| | Left | Right | |
| Overtake slower vehicle | 26 | 61 | 87 |
| Avoid merging vehicle at ramp | 25 | 16 | 41 |
| Exit highway | 6 | 0 | 6 |
| Merge to adjacent lane | 4 | 0 | 4 |
| Avoid parked vehicle on shoulder | 2 | 0 | 2 |
| Other | 1 | 1 | 2 |
| Total | 64 | 78 | 142 |

Fig. 6 shows distributions of the range, $R$, and range rate, $\dot{R}$, for the lane changes to the left and right. The data points are uniformly distributed over the range in the left lane changes, while they are localized in a short distance for the right lane changes due to the motivational difference between the left and right lane changes as mentioned above.

As for the range rate, the signs of the mean values are opposite between the left and right lane changes, i.e., negative (−1.66 m/s) and positive (1.40 m/s), respectively. This sign difference is because the POV is usually faster than the truck in left lane changes and slower in right lane changes with some exceptional cases, e.g., the POV was originally faster but decelerated as the SV changed lanes in a left lane change case, and the POV intended to pass the SV from the right lane in a right lane change case. As a result, the drivers of the large trucks would have to estimate the future gap more carefully in left lane changes than in right lane changes. The dense cluster in Fig. 6(b) may imply that the decision making of changing lanes to the right is trivial since it can be initiated as soon as the large-truck driver has confirmed a positive range.

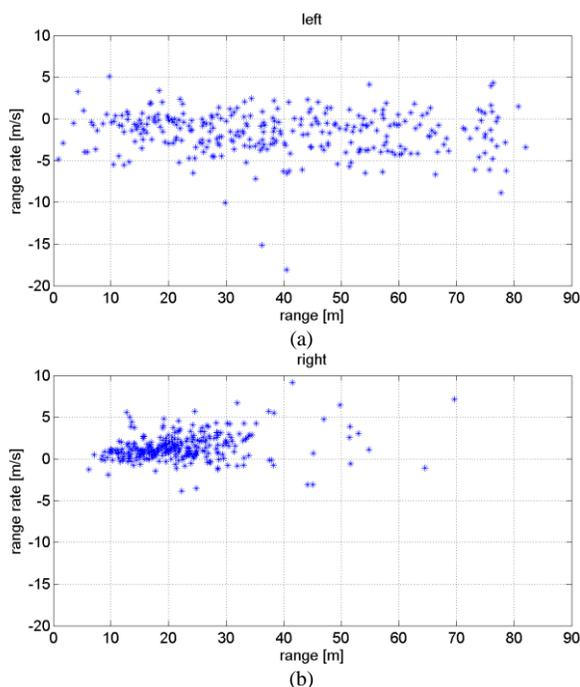

Fig. 6. Range vs. range rate: (a) left lane changes and (b) right lane changes.

Linear regression analyses were conducted to check the correlation between the range and rage rate for both lane change scenarios. The correlation of determination (adjusted-$R^2$) was small for both scenarios, i.e., 0.0039 for left and 0.0614 for right, and the analysis of variance showed that the slope of the regression line for the left lane changes was not statistically significant ($F(1, 298) = 2.17$, $p = 0.142$) but that for the right lane changes was highly statistically significant ($F(1, 298) = 20.6$, $p < 0.0001$). Therefore, no linear correlation between the range and range rate was observed in the left lane changes, suggesting that the closing speed was not linearly related to the distance between the two vehicles during left lane changes. On the other hand, the positive correlation was inferred for the right lane changes with a relatively large dispersion around the regression line, and the small adjusted-$R^2$ value was caused by the small variation in the range, which might be associated with a potential decision criterion that the SV would initiate a lane change as soon as a minimal safe range becomes available regardless of the range rate. In this case, the SV may have gone farther with a larger range rate during the time period between the onset of the maneuver and the lane-change time, which would result in the range rate monotonically increasing as the range at the lane change time as shown in Fig. 6(b).

In general, the SV and/or POV are prone to take a collision avoidance maneuver when the range rate is negative, e.g., the SV accelerates and the POV decelerates, and such maneuvers would be more aggressive for a smaller range or larger negative range rate. Particularly for the SV acceleration, Fig. 7, which shows the relationship between the range at the lane change time and the speed change of the SV in the last 5 s before the lane change time, indicates that the SV tends to accelerate more frequently and faster in case of left lane changes (Fig. 7(a)). On the other hand, in case of right lane changes, the SV tends to stay at the same speed more (Fig. 7(b)) without causing a conflict since it is generally higher than the POV.

The time-to-collision (TTC) [18, 19] is a commonly used conflict metric between two vehicles and is calculated by dividing range by range rate and adding a minus sign,

$$\text{TTC} = -\frac{R}{\dot{R}}. \tag{9}$$

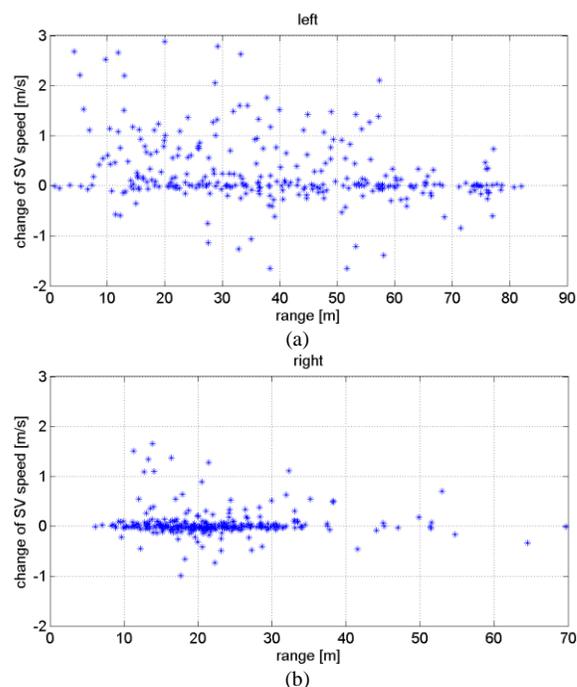

Fig. 7. Change of SV speed within 5 s before the lane change time, plotted against the range at the lane change time: (a) left lane changes and (b) right lane changes.

Thus, a negative TTC (i.e., two vehicles are separating)

indicates that a collision will not happen even if no action is taken by the drivers, so is usually a case that is safe, as long as a sufficiently large range is available so that even if the lead vehicle slows down suddenly the following vehicle can still react to it without a collision. On the other hand, a positive TTC indicates that if neither the POV nor the SV changes speed, a collision is projected to happen. Small positive values of TTC may indicate a potentially risky maneuver, and the smaller the TTC is, the riskier the maneuver may be.

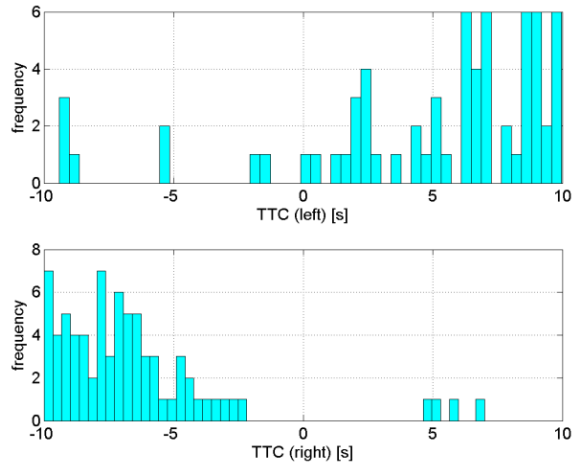

Fig. 8. TTC distributions at the lane change time

The distributions of the TTC between −10 and 10 s are shown in Fig. 8. Left lane changes are riskier than right lane changes by the frequency of positive TTC, and a few events caused small positive values. However, the POV slowing down and/or the SV accelerating were typical in left lane changes, and thus no collisions occurred. On the other hand, there are no notable conflicts in the right lane changes.

Although TTC is a convenient measure because of the simple definition, it may not be suitable for collision avoidance systems since it assumes constant speeds. On the other hand, the acceleration information is expected to provide a more precise predicted conflict measure [19, 20, 20-22]. In this paper, the required deceleration rate for the POV, $D_{req}$, was evaluated with an assumption that the SV was traveling at a constant speed and the POV was to avoid a collision by braking. The expression for $D_{req}$ is given by

$$D_{req} = \frac{\dot{R}^2}{2R} = -\frac{\dot{R}}{2TTC}. \qquad (10)$$

The main advantage of $D_{req}$ is its direct relevance to the severity of the required braking to avoid a rear-end collision, and in fact this is equivalent to the stopping distance model [23] and the constant tau-dot strategy [24, 25]. As shown in Fig. 9, the resulting $D_{req}$ is inversely proportional to TTC but scatters relatively widely in the small TTC region since the effect of range rate in (10) is larger in this region when the denominator is smaller. It is noted that only lane changes with a closing range were considered since potential crashes are of interest.

There is a noticeable difference between the left and right lane changes – 13 left lane changes involved a TTC of less than 4 s, while this was not seen in the right lane changes. The corresponding $D_{req}$ was greater than 0.8 m/s$^2$, which is significantly larger than that in the right lane changes where the maximum $D_{req}$ was about 0.33 m/s$^2$. A video observation indicated that these left lane changes involved collision avoidance maneuvers by the POV, either by braking (6 cases indicated by squares in Fig. 9(a)) or by swerving (7 cases, triangles), rather than merely slowing down by releasing the accelerator. On the other hand, for the right lane changes, the SV imposed little decelerations on the POVs (Fig. 9(b)) and no collision avoidance maneuver was observed. In fact, all collision avoidance maneuvers detected in this study occurred in the left lane changes with TTC < 4 s.

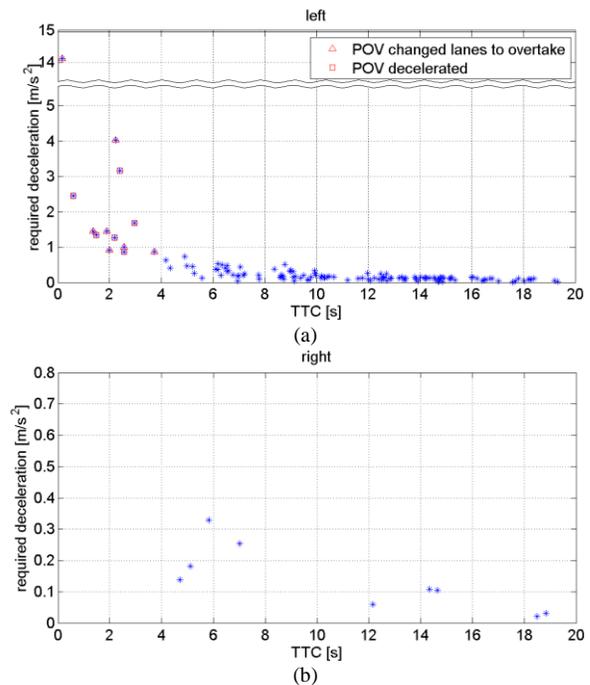

Fig. 9. TTC vs. required deceleration: (a) left lane changes and (b) right lane changes.

From the above results, the following conditions for the warning decision may be suggested: TTC < 4 s or $D_{req}$ > 0.8 m/s$^2$ if the SV is slower than the POV and for right lane changes, particularly with the SV faster than the POV, $R$ < 12.7 m at the 10$^{th}$ percentile value. In practice, the active safety system is required to maintain a sufficient time by taking into account the drivers' reaction time to warning and physical limitations of the vehicle, such as the time required to correct the vehicle path, and thus the system may need to estimate the conflict metrics before the lane change time. In this case, an incorrect assessment of the future driving condition could occur. For example, the system would generate a false positive warning to a predicted unsafe event which is actually a safe event. In order to maximize the overall system performance, it is necessary to balance required design criteria, e.g., to maximize the rate of true positives, minimize the rate of false positives, and maximize the rate of successful countermeasures by any suitable technique, e.g., finding a Pareto set by solving a multi-objective optimization problem. Although this is beyond the scope of this study, it would be interesting to consider a situation where the driver decided *not* to make a lane change – the aforementioned threshold values are solely based on the definition of unsafe situation determined by the forced responses imposed on the POV, and it is unknown what level of conflict would separate viable yet mentally challenging lane changes and the others which could be achieved in comfort.

## V. CONCLUSIONS

This paper presented a study of gap acceptance characteristics of drivers of large trucks in lane change scenarios through the image-based technique with the lane width as the reference measurement using the naturalistic driving data. The major factors affecting the accuracy of the range estimates with the proposed method are the accuracies of the estimated lane width, camera parameters, and locations of the lane markers and POV on images, as well as the image resolution. Accurately locating the lane markers and POV on the image is essential particularly for a far POV, i.e., the estimation accuracy is sensitive to the object distance as the denominator in (1) becomes small, or a slight error in these may cause a large error. In addition, curved roads are more challenging, since road curvature estimates are needed to draw the lane markers on the normalized image plane.

The manual video observations showed that left lane changes typically occur due to a slower vehicle ahead of the truck in the same lane or a vehicle entering the through lane from a ramp. It would be interesting to treat the case of avoiding a merging vehicle independently since the longitudinal distance between the truck and such a vehicle can be very short when it is detected by the truck driver, which is unlikely in the case of overtaking a slower vehicle ahead in the same lane. In this scenario, a mandatory lane change or a dilemma among emergency lane change, hard braking, or acceleration may arise, and this might influence the framework design of the safety system; for example, providing preemptive information about a merging vehicle via the infrastructure and/or that vehicle using wireless communication technologies would enhance the performance of collision avoidance.

The range, range rate, time-to-collision, and required deceleration were obtained by using the range estimation results, and potential threshold values for a warning decision were suggested.

## VI. DISCLAIMERS

The findings and conclusions in the report are those of the authors and do not necessarily represent the views of the National Institute for Occupational Safety and Health (NIOSH). Mention of company names or products does not imply endorsement by NIOSH.

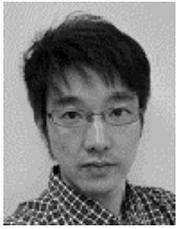

**Kazutoshi Nobukawa** received the B.E. and M.E. degrees in materials science and engineering from Waseda University, Tokyo, Japan, and M.S.E. and Ph.D. degrees in mechanical engineering from University of Michigan, Ann Arbor, MI.

From 2008 to 2010, he was a Graduate Student Research Assistant with the University of Michigan Transportation Research Institute (UMTRI), Ann Arbor, MI. From 2012, He has been a Research Fellow in the UMTRI's Engineering Systems Group. His research interest includes modeling and control of dynamical systems for analyses of collision avoidance systems, vehicle dynamics, tracking, and data mining.

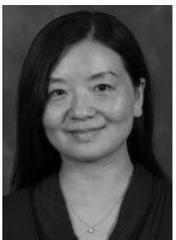

**Shan Bao** received the B.E. and M.E. degree in mechanical engineering from Hefei University of Technology, China and Ph.D. degrees in Industrial engineering from University of Iowa, Iowa City, IA.

Dr. Bao is currently an assistant research scientist in UMTRI's Human Factors Group. She joined UMTRI in 2009, starting as a postdoctoral fellow after completing her Ph.D. in industrial engineering at the University of Iowa. Her research interests focus on driver behavior modeling, driver distraction, naturalistic driving data analysis and driver-simulator study.

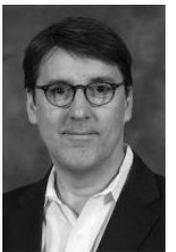

**Dave LeBlanc** received a Ph.D. in aerospace engineering from the University of Michigan, and master's and bachelor's degrees in mechanical engineering from Purdue University.

Dr. David J. LeBlanc is currently an associate research scientist, has been at UMTRI since 1999. Dr. LeBlanc's work focuses on the automatic and human control of motor vehicles, particularly the design and evaluation of driver assistance systems.

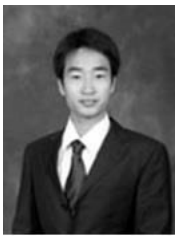

**Ding Zhao** received B.S. degree from Jilin University, China in 2010.

He is currently a Ph.D. student with the Department of Mechanical Engineering, University of Michigan, Ann Arbor, MI, USA. His research interests include vehicle dynamics and control, driver modeling and intelligent transportation.

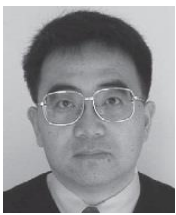

**Huei Peng** received the Ph.D. degree from the University of California, Berkeley, CA, USA, in 1992. He is currently a Professor with the Department of Mechanical Engineering, University of Michigan, Ann Arbor, MI, USA. He is currently the U.S. Director of the Clean Energy Research Center—Clean Vehicle Consortium, which supports 29 research projects related to the development and analysis of clean vehicles in the U.S. and in China. He also leads an education project funded by the Department of Energy to develop ten undergraduate and graduate courses, including three laboratory courses focusing on transportation electrification.

He has more than 200 technical publications, including 85 in refereed journals and transactions. In the last ten years, he was involved in the design of several military and civilian concept vehicles, including Future Tactical Truck Systems, Family of Medium Tactical Vehicles, and Super-High-Mobility Multipurpose Wheeled Vehicle (HUMMWV)—for both electric and hydraulic hybrid-vehicle concept designs. His research interests include adaptive control and optimal control, with emphasis on their applications to vehicular and transportation systems. His current research focuses include design and control of hybrid vehicles and vehicle active safety systems.

Dr. Peng has been an active member of the Society of Automotive Engineers (SAE) and the ASME Dynamic Systems and Control Division (DSCD). From 1995 to 1997, he served as the Chair of the ASME DSCD Transportation Panel. He is a member of the Executive Committee of the ASME DSCD. He served as an Associate Editor for the IEEE/ASME TRANSACTIONS ON MECHATRONICS from 1998 to 2004 and for the ASME Journal of Dynamic Systems, Measurement and Control from 2004 to 2009. He received the National Science Foundation Career award in 1998. He is an ASME Fellow. He is a Changjiang Scholar at Tsinghua University.

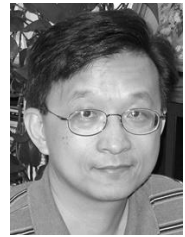

**Christopher S. Pan** is a senior researcher with the National Institute for Occupational Safety and Health (NIOSH), in Morgantown, West Virginia. He received his M.S. in 1989 and Ph.D. in 1991 in Industrial Engineering from the University of Cincinnati and has been conducting research at NIOSH since 1989, with projects chiefly focusing on ergonomics/safety. He currently has an appointment as an adjunct professor in the Department of Industrial and Management Systems Engineering at West Virginia University. He currently serves as a project officer at NIOSH for six funded studies in construction and transportation sectors, including a follow-up collaborative project (2013–2015) of a motor vehicle study with the University of Michigan Transportation Research Institute (UMTRI). For these and related research endeavors, he has been recognized by distinguished peers and professionals in the occupational safety and health community as a competent safety professional, project manager, ergonomist, inventor, and scientist.